\documentclass[a4paper,10pt,showpacs,showkeys,twocolumn,notitlepage]{revtex4-1}
\usepackage{amsmath,amssymb,textcomp,float,dsfont}
\usepackage[utf8]{inputenc}
\usepackage{amsmath}
\usepackage{amsfonts}
\usepackage{amssymb}
\usepackage{times,fullpage}
\usepackage{comment,graphicx}
\usepackage{array}

\graphicspath{{./figures/}}

\begin{document}
\newcommand{\nwc}{\newcommand}
\nwc{\vs}{\vspace}
\nwc{\hs}{\hspace}
\nwc{\la}{\langle}
\nwc{\ra}{\rangle}
\nwc{\nn}{\nonumber}
\nwc{\Ra}{\Rightarrow}
\nwc{\wt}{\widetilde}
\nwc{\lw}{\linewidth}
\nwc{\ft}{\frametitle}
\nwc{\ben}{\begin{enumerate}}
\nwc{\een}{\end{enumerate}}
\nwc{\bit}{\begin{itemize}}
\nwc{\eit}{\end{itemize}}
\nwc{\dg}{\dagger}
\nwc{\mA}{\mathcal A}
\nwc{\mD}{\mathcal D}
\nwc{\mB}{\mathcal B}

\nwc{\Tr}[1]{\underset{#1}{\mbox{Tr}}~}
\nwc{\pd}[2]{\frac{\partial #1}{\partial #2}}
\nwc{\ppd}[2]{\frac{\partial^2 #1}{\partial #2^2}}
\nwc{\fd}[2]{\frac{\delta #1}{\delta #2}}
\nwc{\pr}[2]{K(i_{#1},\alpha_{#1}|i_{#2},\alpha_{#2})}
\nwc{\av}[1]{\left< #1\right>}

\nwc{\zprl}[3]{Phys. Rev. Lett. ~{\bf #1},~#2~(#3)}
\nwc{\zpre}[3]{Phys. Rev. E ~{\bf #1},~#2~(#3)}
\nwc{\zpra}[3]{Phys. Rev. A ~{\bf #1},~#2~(#3)}
\nwc{\zjsm}[3]{J. Stat. Mech. ~{\bf #1},~#2~(#3)}
\nwc{\zepjb}[3]{Eur. Phys. J. B ~{\bf #1},~#2~(#3)}
\nwc{\zrmp}[3]{Rev. Mod. Phys. ~{\bf #1},~#2~(#3)}
\nwc{\zepl}[3]{Europhys. Lett. ~{\bf #1},~#2~(#3)}
\nwc{\zjsp}[3]{J. Stat. Phys. ~{\bf #1},~#2~(#3)}
\nwc{\zptps}[3]{Prog. Theor. Phys. Suppl. ~{\bf #1},~#2~(#3)}
\nwc{\zpt}[3]{Physics Today ~{\bf #1},~#2~(#3)}
\nwc{\zap}[3]{Adv. Phys. ~{\bf #1},~#2~(#3)}
\nwc{\zjpcm}[3]{J. Phys. Condens. Matter ~{\bf #1},~#2~(#3)}
\nwc{\zjpa}[3]{J. Phys. A ~{\bf #1},~#2~(#3)}
\nwc{\zpjp}[3]{Pramana J. Phys. ~{\bf #1},~#2~(#3)}

\title{Landauer bound for erasure using non-ideal gas}
\author{P. S. Pal$^{a,b}$\email{} and A. M. Jayannavar$^{a,b}$\email{}} 
\email{priyo@iopb.res.in, jayan@iopb.res.in}
\affiliation{$^a$Institute of Physics, Sachivalaya Marg, Bhubaneswar-751005, India\\$^b$Homi Bhabha National Institute, Training School Complex, Anushakti Nagar, Mumbai 400085, India}
\begin{abstract}
 Landauer principle states that erasure of $N$ bit information requires an entropic cost of $Nk_B\ln 2$. This fact can easily be demonstrated by relaxation of an ideal gas consisting of $N$ particles
 inside a fixed volume. In this paper we discuss the modification of Landauer bound when we use non-ideal gas with inter-particle interactions for erasure procedure. We have found that the bound 
on the entropy production can be lowered when the interaction between particles is square-well potential. 
\end{abstract}
\pacs{05.30.-d, 03.67.-a, 05.70.-a, 89.70.Cf}
\maketitle

\section{Introduction}
Maxwell's demon \cite{max,vedral_09}, along with Szilard engine\cite{Szilard}, had put forward  a new fundamental concept in the field of information processing - which being a physical quantity 
should obey the laws of thermodynamics.
Information processing involves measurement, information storage and erasure. It was initially thought by researchers that measurement and information writing costs some energy which makes the 
cyclic process in Szilard engine compatible with Second law of thermodynamics. Later in 1961, Landauer proposed that it is not information writing rather information erasure which is accompanied by 
an unavoidable cost in thermodynamic entropy. Landauer principle \cite{landauer_61,landauer_91,barbara} states that the thermodynamic entropy of our surroundings must increase by at least $k_B\ln 2$ 
for each bit of information erased, i.e., while erasing a single bit of information $k_BT \ln 2$ amount of heat is required.

\section{Model}
We consider a system decribed by time dependent Hamiltonian $H(t)$ and is in contact with a heat bath at temperature $T$. The phase space of the system is denoted as $\Gamma$. The state of the
system is described by its probability distribution at any time $\rho(\Gamma,t)$. The average internal energy at any instant of  time is given by
\begin{equation}
 E(t)=\int \rho(\Gamma,t)H(t)d\Gamma.
 \label{ene}
\end{equation}
The non-equilibrium system entropy is defined by the von Neumann entropy\cite{saha,esposito}
\begin{equation}
 S(t)=-\int \rho(\Gamma,t)\ln\rho(\Gamma,t)d\Gamma,
 \label{ent}
\end{equation}
and the corresponding non-equilibrium free energy is given by \cite{esposito}
\begin{equation}
 F(t)=E(t)-TS(t).
 \label{free}
\end{equation}
Now the system undergoes a certain process due to changes in system parameters using some specified protocol. Let $W(t)$ and $Q(t)$ be the work done on the system and the heat transferred from the 
bath to the system at any time $t$. According to the first law of thermodynamics \cite{sekimoto}
\begin{equation}
 \Delta E=W(t)+Q(t).
 \label{1st_law}
\end{equation}
The equilibrium values of internal energy $E^{eq}(t)$, entropy $S^{eq}(t)$ and free energy $F^{eq}(t)$ at any time $t$ can be calculated by considering equilibrium distribution
$\rho(\Gamma,t)=\rho^{eq}(\Gamma,t)=\exp[-\beta\{H(t)-F^{eq}(t)\}]$ and using Eq. \ref{ene}, \ref{ent} and \ref{free}. The free energy of a nonequilibrium state is higher than that of the
corresponding equilibrium state by an amount equal to the temperature times the information $I(t)$ needed to specify the nonequilibrium state:
\begin{eqnarray}
 &&F(t)-F^{eq}(t)\nn\\
 &=&E(t)+T\int \rho(\Gamma,t)\ln\rho(\Gamma,t)d\Gamma-F^{eq}(t)\nn\\
 &&+T\int \rho(\Gamma,t)\ln\rho^{eq}(\Gamma,t)d\Gamma\nn\\
 &&-T\int \rho(\Gamma,t)\ln\rho^{eq}(\Gamma,t)d\Gamma\nn\\
 &=&E(t)+T\int \rho(\Gamma,t)\ln\left[\frac{\rho(\Gamma,t)}{\rho^{eq}(\Gamma, t)}\right] d\Gamma-F^{eq}(t)\nn\\
 &&+\int \rho(\Gamma,t)[-\beta\{H(t)-F^{eq}(t)\}]d\Gamma\nn\\
 &=&TD[\rho(t)||\rho^{eq}(t)]=TI(t).
\end{eqnarray}
The change in  non-equilibrium system entropy $\Delta S$ consists of two contributions namely a reversible part due to heat flow which is called entropy flow $\Delta_e S$ and the other part is 
irreversible and non-negative, which is termed as entropy production $\Delta_i S$
\begin{eqnarray}
 &&\hspace{0.8 cm} \Delta S=\Delta_e S+\Delta_i S,\\
 &&\Delta_e S=\frac{Q(t)}{T};\hspace{0.5 cm}\Delta_i S\geq 0.\nn
\end{eqnarray}
One can combine the above expressions with first law Eq.\ref{1st_law} to obtain an equivalent form of non-equilibrium second law of thermodynamics
\begin{eqnarray}
\Delta E&=&W(t)+Q(t)=W(t)+T\Delta_e S\nn\\
 &=&W(t)+T(\Delta S-\Delta_i S)\nn\\
\Rightarrow T\Delta_i S&=&W(t)-(\Delta E-T\Delta S)\nn\\
 &=&W(t)-\Delta F(t)\geq 0.
\end{eqnarray}
Combination of the above results allows to rewrite the second law under the form of the non-equilibrium Landauer principle
\begin{eqnarray}
 W_{irr}(t)&\equiv& W(t)-\Delta F^{eq}(t)\nn\\
 &=&T\Delta_i S(t)+\Delta F(t)-\Delta F^{eq}(t)\nn\\
 &=&T\Delta_i S(t)+T\Delta I(t).
\label{esp}
\end{eqnarray}
Here $\Delta I(t)=I(t)-I(0)$. $I$ can be indentified with the amount of information that is required to be processed to switch from known equilibrium distribution $\rho^{eq}(t)$ to the distribution 
$\rho(t)$.For a relaxation process $W_{irr}(t)=0$ and $I(t)=0$ as  the system ends up in an equilibrium state. Hence, from Eq.\ref{esp} it can be easily shown that  the information gained at the 
starting of the process i.e., $I(0)$ is completely lost into entropy production $\Delta_i S(t)$.

A non-ideal gas consisting of $N$ particles is confined in one half of a box of length $L$, surface area $A$ and hence volume $V=AL$. The system is in contact with a heat bath at inverse temperature 
$\beta$ and it is at equilibrium. This is the initial condition of the sytem. Now at time $t=0$, we instantly remove the partition that was dividing the box into two halves and allow the gas to relax 
within the whole box. No work will  be done by the gas. The information $I$ stored in the system will convert into irreversible entropy production. We want to calculate $I$.

\begin{figure}[H]
 \begin{center}
  \includegraphics[height=2 in,width=3 in]{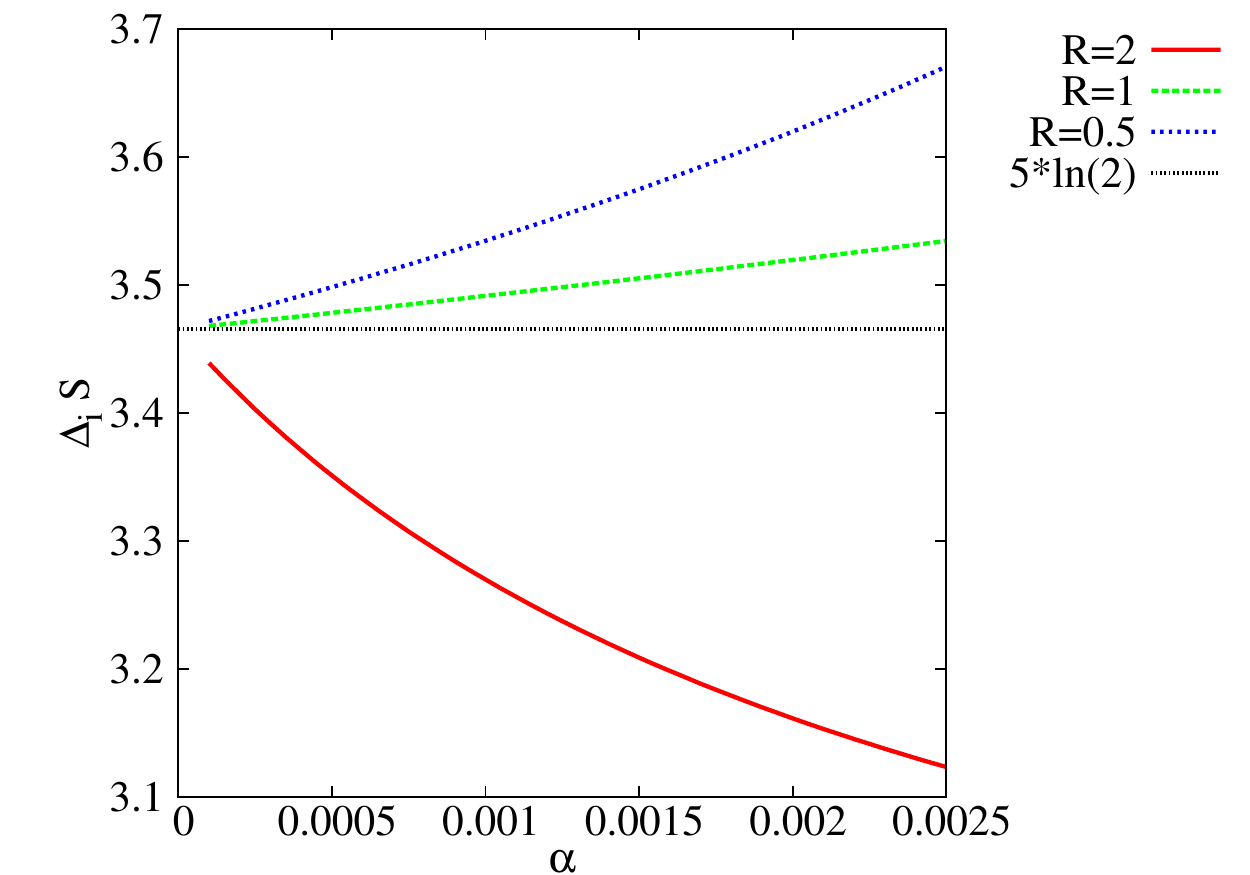}
  \caption{ Entropy production as a function $\alpha$ for square well potential with $N=5$. The well depth $\varepsilon$ and the temperature is taken to be unity.}
  \label{ent_vs_alpha}
 \end{center}
\end{figure}

\begin{figure}[H]
 \begin{center}
  \includegraphics[height=2 in,width=3 in]{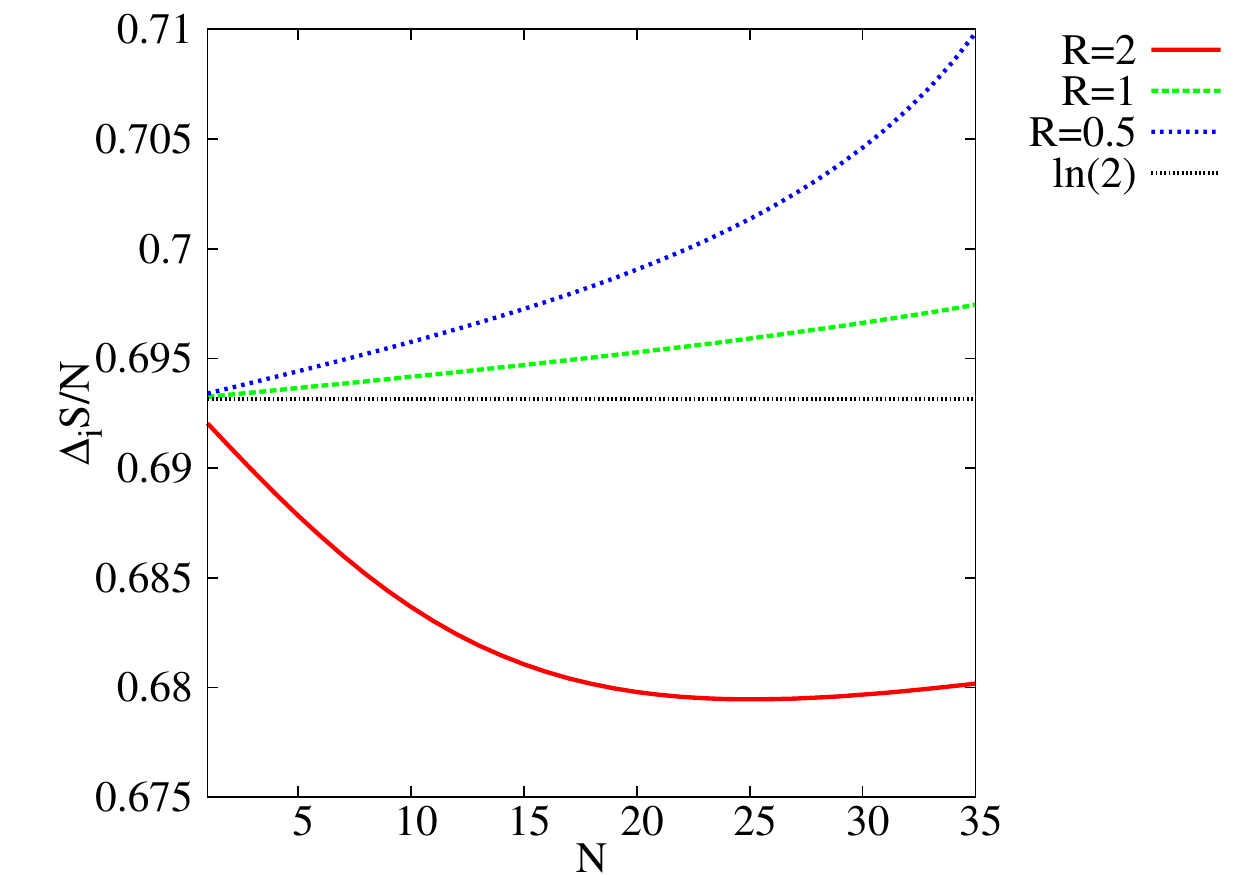}
  \caption{ Entropy production per particle as a function number of particles for square well potential with different well width $R$. The well depth $\varepsilon$ and the temperature
  is taken to be unity. $\alpha$ is taken to be $0.0001$.}
  \label{ent_vs_N}
 \end{center}
\end{figure}

\begin{figure}[H]
 \begin{center}
  \includegraphics[height=2 in,width=3 in]{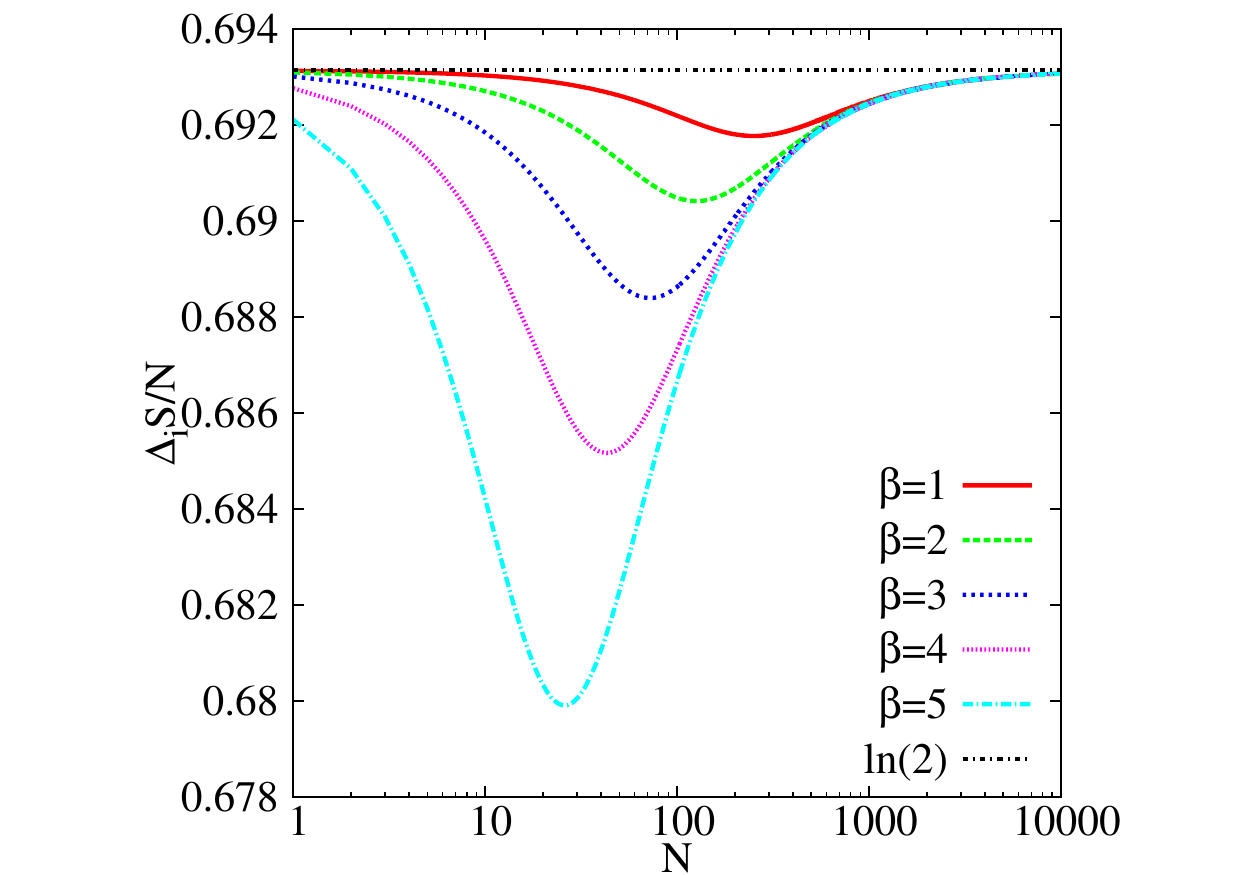}
  \caption{ Entropy production per particle as a function number of particles in case of square well potential for different temperatures. The well depth $\varepsilon$ is considered to be unity and 
  $\alpha$ is taken to be $0.000001$.}
  \label{ent_vs_N_R2}
 \end{center}
\end{figure}

If $\Gamma$ represents the phase space co-ordinates of $N$ particles and $H$ is the Hamiltonian of the $N$ particle system, then the probability density of the system at $t=0^+$ is given by
\begin{eqnarray}
 \rho(\Gamma)&=&\frac{e^{-\beta H(\Gamma)}}{Z(V/2)} \hspace{1 cm} 0\leq \{x_i\}_{i=1}^N\leq L/2\nn\\
 &=& 0 \hspace{2 cm} L/2\leq \{x_i\}_{i=1}^N\leq L,
\end{eqnarray}
where $x_i$ is the position co-ordinate  of the $i$-th  particle. After relaxation, the system will settle to an equilibrium state corresponding to the entire box and its state will be given by 
\begin{equation}
 \rho^{eq}(\Gamma)=\frac{e^{-\beta H(\Gamma)}}{Z(V)},
\end{equation}
where $Z(V)$ is the partition function of $N$-particle non-ideal gas given by \cite{net}
\begin{equation}
 Z(V)=\frac{cL^N}{N!}\left[1-\frac{N^2}{V}B_2(\beta)\right].
 \label{pf}
\end{equation}
Here $c=(1/N!)(V/\Lambda^3)^N$  and $B_2(\beta)$ is the second virial co-efficient which depends on the interaction between the particles of the gas. Eq.\ref{pf} has been derived 
with low density approximation. Now the information stored in the system is given by 
\begin{eqnarray}
 I&=&\int \rho(\Gamma)\ln\left[\frac{\rho(\Gamma)}{\rho^{eq}(\Gamma)}\right]d\Gamma,\nn\\
 &=&\int \rho(\Gamma)\ln\left[\frac{Z(V)}{Z(V/2)}\right]d\Gamma,\nn\\
 &=& \ln\left[\frac{Z(V)}{Z(V/2)}\right],\nn\\
 &=&N\ln 2+\ln\left[\frac{1-(N^2/V)B_2(\beta)}{1-(2N^2/V)B_2(\beta)}\right]
\end{eqnarray}
\begin{figure}[H]
 \begin{center}
  \includegraphics[height=2.5 in,width=2.5 in]{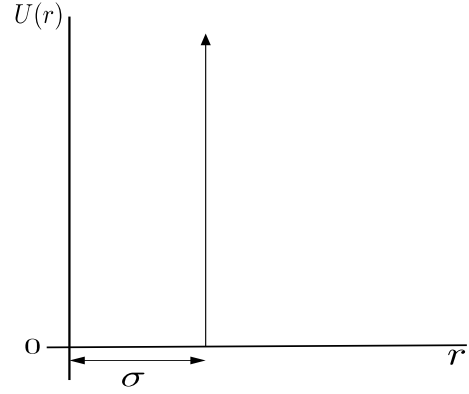}
  \caption{ Hard core potential as a function of interatomic distance $r$. $\sigma$ is the hard core radius.}
  \label{hard_core}
 \end{center}
\end{figure}
In case of hard-core interactions between particles, $B_2(\beta)=(2/3)\pi\sigma^3$, where $\sigma$ is the hard core radius and the information is given by
\begin{equation}
 I_{hc}=N\ln 2+\ln\left[\frac{1-\alpha N^2}{1-2\alpha N^2}\right],
\end{equation}
where $\alpha=(2/3V)\pi\sigma^3$. Since $\alpha N^2$ is always greater than 0, $I_{hc}>N\ln 2$. 
\begin{figure}[H]
 \begin{center}
  \includegraphics[height=2 in,width=3 in]{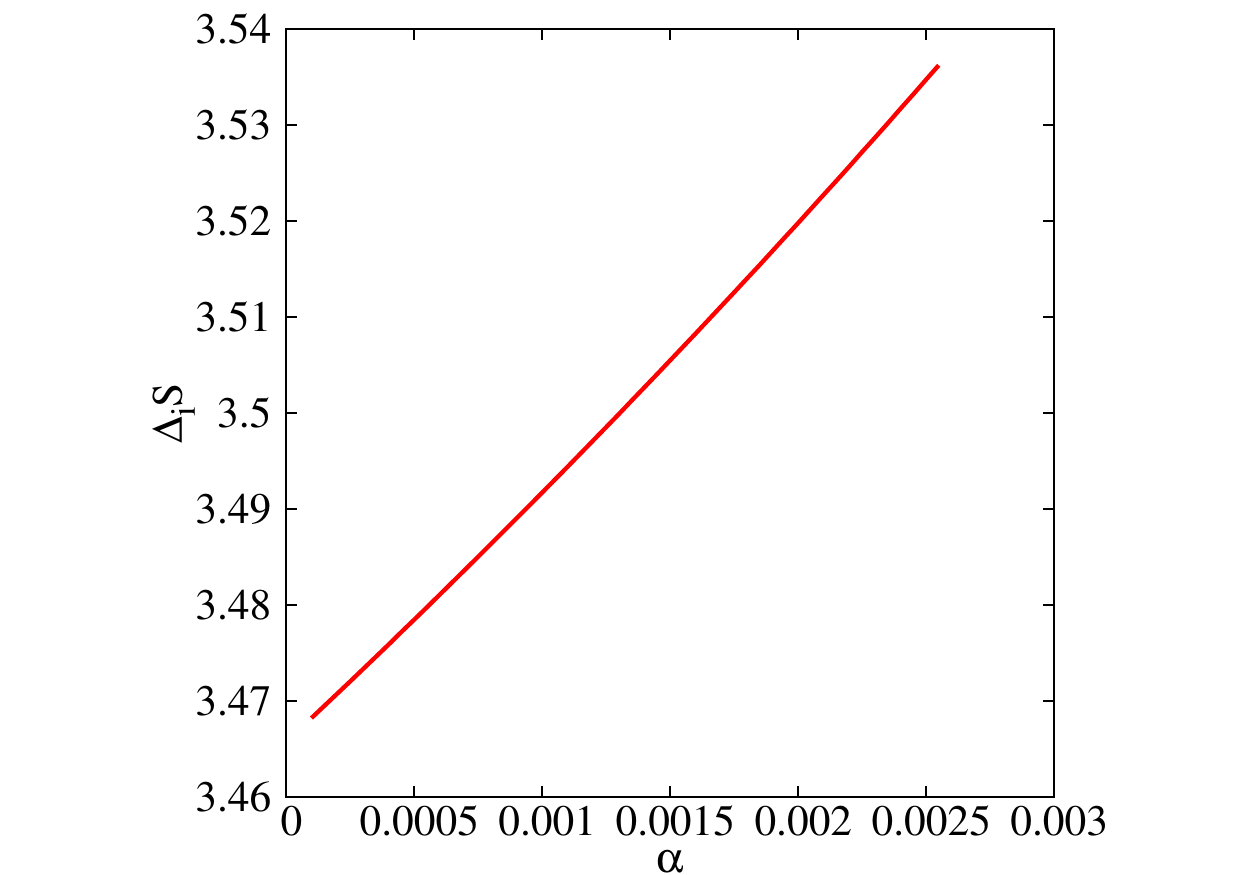}
  \caption{ Entropy production as a function $\alpha$ for hard core potential with $N=5$. For five particles the Landauer bound is $3.465$ which is clearly below the values shown in the plot}
  \label{ent_vs_alp}
 \end{center}
\end{figure}
In Fig.\ref{ent_vs_alp}, the amount of entropy produced due to  information erasure is plotted as a function of $\alpha$ for hard core potential. Entropy production increases with $\alpha$ and the plot
clearly confirms the Landauer principle that erasure of each bit of information is accompanied by releasing of atleast $k_BT\ln 2$ amount of heat.
 \begin{figure}[H]
 \begin{center}
  \includegraphics[height=2.5 in,width=2.5 in]{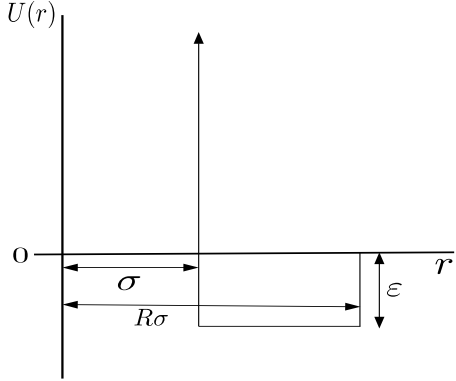}
  \caption{ Square well interaction as function interatomic distance $r$. $\sigma$ is the hard core radius. It is followed by an attractive well of depth $\varepsilon$ and has a width of $R$ 
  times the hard core radius.}
  \label{square_well}
 \end{center}
\end{figure}
In case of square well interaction between particles, $B_2(\beta)=C_1-C_2e^{\beta\varepsilon}$ with $C_1=(2/3)\pi\sigma^3R^3$ and $C_2=(2/3)\pi\sigma^3(R^3-1)$. The stored information is given by 
\begin{equation}
 I_{sw}=N\ln 2+\ln\left[\frac{1-\delta N^2}{1-2\delta N^2}\right],
\end{equation}
where $\delta=(1/V)B_2(\beta)=\alpha R^3-\alpha(R^3-1)e^{\beta\varepsilon}$. $\delta$ can be positive or negative depending on  different parameters. If $\delta<0$, $I_{sw}<N\ln 2$ and 
vice-versa. In fact, $\delta<0$ if $R>1$ and consequently the entropy production due to information erasure goes below Landauer bound. This fact is depicted both in Fig. \ref{ent_vs_alpha} and
\ref{ent_vs_N}. In Fig. \ref{ent_vs_alpha} entropy production is plotted as a function of $\alpha$. When the well width is larger than 1 the entropy production decreases with increasing $\alpha$.
On the other hand for $R\leq 1$, it increases along with $\alpha$.
 In Fig. \ref{ent_vs_N} entropy production per particle is plotted as a function number of particles. When $R\leq 1$, entropy production increases with $N$. 
 For $R>1$, the entropy production is below Landauer bound and it decreases with $N$ upto a certain value and then again increases. 
Entropy production  finally settles to the Landauer bound at  large values of $N$ as shown in Fig.\ref{ent_vs_N_R2}. In case of the last plot we have taken the value of $\alpha$ to be 
very small to stick with the low density approximation.

\section{Conclusions}
In conclusion, we have investigated the Landauer bound with relaxation of non-ideal gas with inter-particle interactions namely hard core and square well potentials. We have found that the bound 
on the entropy production and hence the heat dissipation can be lowered when the interaction between particles is square-well potential. 
\section{Acknowledgement}
 AMJ thanks DST, India for financial support (through J. C. Bose National Fellowship).

\end{document}